\documentclass{iopart_mod}

\usepackage{amsmath}
\usepackage{amsfonts}
\usepackage{mathrsfs}
\usepackage{array}
\usepackage{epsfig}
\usepackage{cite}

\newcommand\MM{\mathcal{M}}

\newcommand\dM{\partial \MM}

\newcommand\FF{\mathcal{F}}
\newcommand\LL{\mathcal{L}}

\newcommand\ZZ{\mathcal{Z}}

\newcommand\nts{\!}
\newcommand\bns{\nts \nts \nts}

\newcommand\de{\delta}

\newcommand\defeq{\mathrel{\mathop:}=}

\DeclareMathAlphabet{\mathpzc}{OT1}{pzc}{m}{it}

\newcommand\CurlD{\mathscr{D}}

\newcommand{\arcsinh}{{\rm arcsinh}\,}

\newcommand{\eq}[2]{\begin{equation} #1 \label{#2} \end{equation}}

\newcommand{\blist}{\begin{itemize}}
\newcommand{\elist}{\end{itemize}}

\begin{document}

\hfill MIT--CTP--3887

\hfill MPP--2007--153

\title{Black Hole Thermodynamics and Hamilton-Jacobi Counterterm}

\author{Luzi Bergamin}

\address{ESA Advanced Concepts Team, ESTEC, Keplerlaan 1,
2201 AZ Noordwijk, The Netherlands}
\ead{Luzi.Bergamin@esa.int}

\author{Daniel Grumiller}

\address{Massachusetts Institute of Technology,
77 Massachusetts Ave, Cambridge, MA 02139, USA}
\ead{grumil@lns.mit.edu}

\author{Robert McNees}

\address{Perimeter Institute for Theoretical Physics,
31 Caroline Street North, Waterloo, Ontario N2L 2Y5, Canada}
\ead{rmcnees@perimeterinstitute.ca}

\author{Ren\'e Meyer}

\address{Max Planck Institut f\"ur Physik, F\"ohringer Ring 6, 80805 M\"unchen, Germany}
\ead{meyer@mppmu.mpg.de}

\begin{abstract}
We review the construction of the universal Hamilton-Jacobi counterterm for dilaton gravity in two dimensions, derive the corresponding result in the Cartan formulation and elaborate further upon black hole thermodynamics and semi-classical corrections. Applications include spherically symmetric black holes in arbitrary dimensions with Minkowski- or $AdS$-asymptotics, the BTZ black hole and black holes in two-dimensional string theory. 
\end{abstract}

%\pacs{}
%\maketitle

\section{Introduction}

There are numerous applications in physics where an action
\eq{
I_{\rm bulk}[\phi]=\int_{\MM}\nts\nts d^nx \,\LL_{\rm bulk}(\phi,\nabla\phi)
}{eq:lp1}
has to be supplemented by boundary terms
\eq{
I_{\rm tot}[\phi]=I_{\rm bulk}[\phi]+\int_{\dM}\nts\nts\nts d^{n-1}x\,\LL_{\rm boundary}(\phi,\nabla_\bot\phi,\nabla_\parallel\phi)\,.
}{eq:lp2}
Here $\nabla_\bot$ and $\nabla_\parallel$ denote the normal and parallel components of the derivative with respect to the boundary $\dM$.
The simplest example is quantum mechanics, where
\eq{
I_{\rm bulk}[q,p] = \int_{t_0}^{t_1}\nts\nts dt\,\left[-q\dot{p}-H(q,p)\right]
}{eq:lp3}
has to be supplemented by a  boundary term
\eq{
I_{\rm tot}[q,p]=I_{\rm bulk}[q,p]+qp|_{t_0}^{t_1}
}{eq:lp4}
if Dirichlet boundary conditions are imposed on the coordinate, $\delta q|_{t_i}=0$. In addition to this ``Gibbons-Hawking-York'' boundary term one can add another boundary term
\eq{
\Gamma[q,p] = I_{\rm tot}[q,p] - \FF(t,q)|_{t_0}^{t_1}
}{eq:}
which depends only on quantities held fixed at the boundary. This seems to be a superfluous addition, as it does neither change the equations of motion nor the variational principle (as opposed to the ``Gibbons-Hawking-York'' boundary term), but in some applications such a term is crucial and determined almost uniquely from consistency requirements: symmetries and accessibility of the classical approximation. One such application is the Euclidean path integral for black holes (BHs), which provides a convenient shortcut to BH thermodynamics.

\section{Hamilton-Jacobi counterterm in two-dimensional gravity}\label{se:2}

The essential features and difficulties arise already in low dimensions. For transparency we focus on two-dimensional (2D) models. The bulk action for 2D dilaton gravity \cite{Grumiller:2002nm},
\begin{equation}\label{eq:bulkaction}
  I_{\rm bulk}[g,X] = - \frac{1}{16\pi G_2}\,\int_{\MM} \nts \nts d^{\,2}x \,\sqrt{g}\, \left[ X\,R - U(X)\,
		\left(\nabla X\right)^2 - 2 \, V(X) \raisebox{12pt}{~}\right] 
\end{equation}
has to be supplemented by a Gibbons-Hawking-York boundary term
\eq{
I_{\rm tot}[g,X] = I_{\rm bulk}[g,X] -  \frac{1}{8\pi G_2}\, \int_{\dM} \bns dx \, \sqrt{\gamma}\,X\,K ~,
}{eq:GHYaction}
if Dirichlet boundary conditions are imposed on the dilaton field $X$ and the induced metric at the boundary $\gamma$. The meaning of all symbols is standard and our notation is consistent with \cite{Grumiller:2007ju}. In addition one could add another boundary term
\eq{
\Gamma[g,X]=I_{\rm tot}[g,X] - \underbrace{\frac{1}{8\pi G_2}\, \int_{\dM} \bns dx \, \sqrt{\gamma}\,\FF(X,\nabla_\parallel X,\gamma,\nabla_\parallel\gamma)}_{I_{\rm CT}[\gamma,X]}\,.
}{eq:angelinajolie}
We demonstrate now why such a term is needed and show that it is determined essentially uniquely from consistency requirements: symmetries and accessibility of the classical approximation.

\subsection{Symmetries}
\label{sec:symmetries}

Diffeomorphism covariance along the boundary requires that $\FF$ in \eqref{eq:angelinajolie} transforms as a scalar. Since there are no scalar invariants constructed from $\gamma$ in one dimension, $\mathcal F$ can be reduced to
%Diffeomorphism covariance with respect to diffeomorphisms along the boundary requires that $\FF$ in \eqref{eq:angelinajolie} transforms as a scalar. Since there are no scalar invariants one could construct from $\gamma$ in one dimension, we know that actually 
%\eq{
$\FF=\FF(X,\nabla_\parallel X)$. %\,.
%}{eq:lp5}
Another simplification arises if we restrict ourselves to isosurfaces of the dilaton field, which is sufficient for our purposes. Then $X$ is constant along the boundary, $\nabla_\parallel X=0$, so that we are left with a function
\eq{
\FF=\FF(X)\,.
}{eq:lp6}
Symmetry requirements have reduced the dependence on four variables in \eqref{eq:angelinajolie} to a dependence on only one variable, $X$.

\subsection{Accessibility of the classical approximation}
\label{sec:accessibility}
While symmetries help to reduce the ambiguities in $\FF$ they do not explain why such a term is needed in the first place. To this end we consider the Euclidean path integral,
\begin{equation}\label{PartitionFunction}
  \ZZ = \int \CurlD g \CurlD X \, \exp\Big(-\frac{1}{\hbar}\,I[g,X] \Big) ~.
\end{equation}
The path integral is evaluated by imposing boundary conditions on the fields and then performing the weighted sum over all relevant space-times $(\MM,g)$ and dilaton configurations $X$. In the classical limit it is dominated by contributions from stationary points of the action. This can be verified by expanding it around a classical solution
\begin{equation}\label{Saddle}
  I[g_{cl} + \delta g, X_{cl} + \delta X] = I[g_{cl},X_{cl}] + \de I[g_{cl}, X_{cl}; \delta g, \delta X] + \frac12 \de^2I[g_{cl}, X_{cl}; \delta g, \delta X] + \ldots	
\end{equation}
where $\de I$ and $\de^2 I$ are the linear and quadratic terms in the Taylor expansion. The saddle point approximation of the path integral 
\begin{equation}\label{ApproxPF}
 \ZZ \sim \exp\Big(-\frac{1}{\hbar}\,I[g_{cl},X_{cl}]\Big) \, \int \CurlD \delta g \CurlD \delta X \, \exp\Big(-\frac{1}{2\hbar}\,\de^2I[g_{cl}, X_{cl}; \delta g,\delta X] \Big) 
\end{equation}
is defined if:
\begin{enumerate}
\item The on-shell action is bounded from below, $I[g_{cl},X_{cl}]>-\infty$.
\item The first variation vanishes on-shell, $\de I[g_{cl}, X_{cl}; \delta g, \delta X] =0$ for all variations $\delta g$ and $\delta X$ preserving the boundary conditions.
\item The second variation has the correct sign for convergence of the Gaussian in \eqref{ApproxPF}.
\end{enumerate}
The last condition actually means consistency of the semi-classical approximation, and we shall not discuss it here. Instead, we focus on the first two conditions. Both are violated for typical BH solutions of \eqref{eq:GHYaction} if the boundary is located in the asymptotic region $X\to\infty$:
\begin{enumerate}
\item The on-shell action behaves as $I_{\rm tot}[g_{cl},X_{cl}\to\infty]=2M/T-S-w(X_{cl}\to\infty)/T$, and $\lim_{X\to\infty} w(X)\to\infty$ for most models of interest.\footnote{$T$ is the Hawking temperature and $S$ the Bekenstein-Hawking entropy. Both are determined from the mass $M$ and the functions $Q(X)$ and $w(X)$ defined in the Appendix.}
\item The first variation of the action receives a boundary contribution \begin{equation}\label{ActionVariationIntro}
   \left.\delta I_{\rm tot} \raisebox{12pt}{\,}\right|_{\rm on-shell} \sim \int_{\dM} \bns dx \, \sqrt{\gamma}\,
	\left[ \pi^{ab} \, \delta \gamma_{ab} + \pi_{X} \, \delta X \, \right] \neq 0 
\end{equation}
because the product $\pi^{ab} \, \delta \gamma_{ab}$ is non-vanishing: the variation of the induced metric $\gamma_{ab}$ does not fall off sufficiently fast to compensate for the divergence of the momenta $\pi^{ab}$.
\end{enumerate}
We emphasize that an ad-hoc subtraction $I_{\rm ren}[g_{cl},X_{cl}\to\infty]:=I_{\rm tot}[g_{cl},X_{cl}\to\infty]+w(X_{cl}\to\infty)/T$ is inconsistent: while it leads to a finite on-shell action it does not address the second problem. Both can be solved by choosing $\FF(X)$ adequately.
%, which however can be solved by choosing $\FF(X)$ adequately. In this way, also the first law of thermodynamics will be fulfilled automatically.

Since the on-shell action solves the Hamilton-Jacobi equation one can expect cancellations if also $\FF$ is a solution to the Hamilton-Jacobi equation. Thus, our guiding principle is to demand that $\FF$ be a solution of the Hamilton-Jacobi equation (see the next Section for details). 
%, which in our case turns into a first order ordinary differential equation. 
This method was applied first to the Witten BH and to type 0A string theory \cite{Davis:2004xi} and later generalized to generic 2D dilaton gravity \cite{Grumiller:2007ju}. The result is
\begin{equation}\label{LCT}
  \FF(X) =  - \sqrt{\left(w(X) + c\right)e^{-Q(X)}}
\end{equation}
where $c$ is an integration constant. It can be absorbed into a redefinition of $w$ (cf.~Appendix) and reflects the freedom to choose the ground state of the system. Thus, without loss of generality we can set it to zero and finally obtain a consistent action \cite{Grumiller:2007ju}
\eq{
\begin{split}
  \Gamma[g,X] &= - \frac{1}{16\pi G_2}\,\int_{\MM} \nts \nts d^{\,2}x \,\sqrt{g}\, \left[ X\,R - U(X)\,
		\left(\nabla X\right)^2 - 2 \, V(X) \right] \\ 
&\quad - \frac{1}{8\pi G_2}\, \int_{\dM} \bns dx \, \sqrt{\gamma}\,X\,K 
+ \frac{1}{8\pi G_2}\, \int_{\dM} \bns dx \,\sqrt{\gamma}  \,  
	\sqrt{w(X) \, e^{-Q(X)}} \,.
\end{split}
}{ActionConclusion}
The classical approximation is now well-defined because
\begin{enumerate}
\item $\Gamma[g_{cl},X_{cl}\to\infty]=M/T-S$ is finite.
\item $\delta\Gamma[g_{cl},X_{cl};\delta g,\delta X]=0$ for all $\delta g$ and $\delta X$ preserving the boundary conditions.
\end{enumerate}
Moreover, as opposed to \eqref{eq:GHYaction} the action \eqref{ActionConclusion} is consistent with the first law of thermodynamics. Perhaps the most remarkable property of the counterterm in \eqref{ActionConclusion} is its universality: while usually different subtraction methods are employed depending on whether spacetime is asymptotically flat, $AdS$ or neither of both, our result is not sensitive to the asymptotics. This universality does not appear to exist in higher dimensions or even in 2D if `standard' subtraction methods are used \cite{Regge:1974zd}. % , Gibbons:1976ue,Liebl:1997ti, Henningson:1998gx, Balasubramanian:1999re, Kraus:1999di, Emparan:1999pm, deHaro:2000xn, Papadimitriou:2005ii, McNees:2005my, Mann:2005yr, Mann:2006bd}.

\section{Cartan formulation}
In many applications a first order formulation in terms of Cartan variables is advantageous \cite{Grumiller:2002nm}. Therefore we now derive the Hamilton-Jacobi counterterm in this formulation. The corresponding action (we set $8\pi G_2=1$)
\begin{equation}
\label{FOaction}
 \begin{split}
 I^{\rm FO}_{\rm tot}[X, Y^a, e_a, \omega] &= - \int_{\mathcal M} \left[ Y^a D e_a + Xd\omega + \epsilon\,(\frac12 U(X)Y^aY_a+V(X)) \right] \\ &\quad + \int_{\partial \mathcal M} \left[ X\omega_{\|} + \frac{i}{2} Xd\ln \frac{e_{\|}}{\bar{e}_{\|}}\right]\ ,
\end{split}
\end{equation}
contains the Cartan variables $\omega$ and $e_a$, as well as the scalar fields $X$ and $Y^a$ (we use a complexified dyad, $\bar{e}=e^\ast$, cf.~\cite{Bergamin:2004pn} for the details of our notation). As \eqref{FOaction} is classically equivalent to \eqref{eq:GHYaction} (cf.~the Appendix of Ref.~\cite{Bergamin:2004pn}) it will suffer from the same problems as described in Section \ref{sec:accessibility}. We follow the same strategy as in the second order formulation \cite{Davis:2004xi,Grumiller:2007ju} to find the corresponding Hamilton-Jacobi counterterm $I_{\rm CT}^{\rm FO}$, which, using the arguments in Section \ref{sec:symmetries}, can be reduced to  
\begin{equation}
\label{FOCT}
\Gamma^{\rm FO}[X, Y^a, e_a, \omega] =  I^{\rm FO}_{\rm tot}[X, Y^a, e_a, \omega] - \underbrace{\int_{\dM} \bns dx^{\|} \, \sqrt{2e_{\|} \bar e_{\|}}\,\FF(X)}_{I_{\rm CT}^{\rm FO}[X,e_{\|}\bar{e}_{\|}]}\, .
\end{equation}
%For concreteness the boundary is chosen at fixed $x^1$, while $x^0$ plays the role of time in a Hamiltonian analysis. 
The variation of the action produces the equations of motion plus the boundary term\footnote{Contributions emerging from the logarithm in the boundary term are dropped as we assume that the boundary is an isosurface of the dilaton and that there is no boundary of the boundary. The coordinate along the boundary, $x^{\|}$, can be thought of as Euclidean time $\tau$.}
\begin{equation}
 \int_{\partial \mathcal M} \bns d x^{\|} \,\left[ Y \delta \bar{e}_{\|} + \bar{Y} \delta \bar e_{\|} - \delta X \omega_{\|}\right]\, .
\end{equation}
To cancel it we assign Dirichlet boundary conditions to $X$, $e_{\|}$ and $\bar{e}_{\|}$. As in \cite{Davis:2004xi,Grumiller:2007ju} it is possible to write the momenta which are not fixed at the boundary,
\begin{align}
\label{OnShellVar}
 \omega_{\|} &= - \left.\frac{\delta I^{\rm FO}_{\rm tot}}{\delta X}\right|_{\rm on-shell} \,, & Y &= \left.\frac{\delta I^{\rm FO}_{\rm tot}}{\delta \bar{e}_{\|}}\right|_{\rm on-shell} \,,  & \bar{Y} &= \left.\frac{\delta I^{\rm FO}_{\rm tot}}{\delta e_{\|}} \right|_{\rm on-shell} \,,
\end{align}
as variations of the on-shell action. The Hamilton constraint
\eq{
- \omega_{\|} \frac{Y\bar{e}_{\|}+\bar{Y}e_{\|}}{2e_{\|}\bar{e}_{\|}} + U(X) Y\bar{Y}+ V(X) = 0\,,
}{eq:hc}
follows from a standard constraint analysis of the first order action \eqref{FOaction}. By construction the Hamilton-Jacobi counterterm must be a solution of this constraint. Replacing in \eqref{OnShellVar} the on-shell action $I_{\rm tot}^{\rm FO}$ by the counterterm $I_{\rm CT}^{\rm FO}$, plugging this into the Hamilton constraint \eqref{eq:hc} and exploiting that $I_{\rm CT}^{\rm FO}$ depends solely on the combination $e_{\|}\bar{e}_{\|}$ establishes
\eq{
\frac{\delta I_{\rm CT}^{\rm FO}}{\delta X} \frac{\delta I_{\rm CT}^{\rm FO}}{\delta(e_{\|}\bar{e}_{\|})} + U(X) e_{\|}\bar{e}_{\|} \left(\frac{\delta I_{\rm CT}^{\rm FO}}{\delta(e_{\|}\bar{e}_{\|})}\right)^2 + V(X) = 0\,.
}{eq:FDEFOWTF}
This functional differential equation for the counterterm by virtue of the Ansatz \eqref{FOCT} simplifies to
\eq{
\frac{d}{dX} \FF^2(X) + U(X) \FF^2(X) + 2V(X) = 0\,.
}{eq:ODEFFO}
The solution of this first order ordinary differential equation is given by\footnote{There is a sign ambiguity since \eqref{eq:ODEFFO} yields only $\FF^2(X)$. The sign choice in \eqref{eq:FFO} gives an action with a consistent classical limit.}
\eq{
\FF(X) =  - \sqrt{\left(w(X) + c\right)e^{-Q(X)}}\,.
}{eq:FFO}
This coincides with \eqref{LCT} and thus we conclude that the Hamilton-Jacobi counterterms in second and first order formalisms are identical, as might have been anticipated on general grounds. Setting again $c=0$, the consistent first order action is
\begin{equation}
\label{GammaFO}
 \begin{split}
 & \Gamma^{\rm FO}[X, Y^a, e_a, \omega] = - \int_{\mathcal M} \left[ Y^a D e_a + Xd\omega + \epsilon\,(\frac12 U(X)Y^aY_a+V(X)) \right] \\ &\quad + \int_{\partial \mathcal M} \left[ X\omega_{\|} + \frac{i}{2} Xd\ln \frac{e_{\|}}{\bar{e}_{\|}}\right] + \int_{\dM} \bns dx^{\|} \, \sqrt{2e_{\|} \bar e_{\|}}\,\sqrt{w(X)e^{-Q(X)}}\, . 
\end{split}
\end{equation}

\section{Black hole thermodynamics and further applications}

An immediate consequence of our result \eqref{ActionConclusion} is the Helmholtz free energy \cite{Grumiller:2007ju}
\begin{equation}\label{FreeEnergy}
  F_c(T_c,X_c) = \frac{1}{8\pi G_2}\,\sqrt{w(X_c) e^{-Q(X_c)}}\left(1-\sqrt{1-\frac{2M}{w(X_c)}}\right)  - \frac{X_h}{4 G_2} \, T_c \,,
\end{equation} 
which is related to the on-shell action in the usual way, $F_c= T_c \, \Gamma_c$. Here $X_c$ denotes the value of the dilaton field at the location of a cavity wall in contact with a thermal reservoir, while $X_h$ denotes the value of the dilaton at the BH horizon. The local temperature $T_c$ is related to the Hawking temperature $T$ by the standard Tolman factor, $T_c=T/\sqrt{\xi(X)}$. All other quantities are defined in the Appendix. The entropy,
\eq{
S=-\left.\frac{\partial F_c}{\partial T_c}\right|_{X_c} = \frac{X_h}{4 G_2} = \frac{A_h}{4 G_{\rm eff}}\,,
}{eq:S}  
is in agreement with the Bekenstein-Hawking result. Here $A_h=1$ because we are in 2D, and $G_{\rm eff}=G_2/X_h$. For dimensionally reduced models \eqref{eq:S} can be interpreted also from a higher-dimensional perspective: $A_h\propto X_h$ and $G_{\rm eff}\propto G_2$, with the same proportionality constants. The result \eqref{eq:S} is well-known and was obtained by various methods \cite{Gibbons:1992rh}. %, Nappi:1992as,Frolov:1992xx,Gegenberg:1994pv}. 
However, the free energy \eqref{FreeEnergy} contains a lot of additional information and allows a quasi-local treatment of BH thermodynamics (where applicable in agreement with \cite{Brown:1991gb}), %,Brown:1992br}), 
including stability considerations. For an extensive study of thermodynamical properties and more Refs.~we refer to \cite{Grumiller:2007ju}. 

The class of BHs described by the action \eqref{ActionConclusion} or \eqref{GammaFO} is surprisingly rich (cf.~e.g.~table 1 in \cite{Grumiller:2006rc}), and includes spherically symmetric BHs (like Schwarzschild or Schwarzschild-$AdS$) in any dimension, spinning BHs in three dimensions \cite{Banados:1992wn} %,Banados:1992gq} 
and string BHs in two dimensions \cite{Elitzur:1991cb, %Mandal:1991tz,Witten:1991yr,
Dijkgraaf:1992ba}. As an example we consider now the exact string BH \cite{Dijkgraaf:1992ba}, and review some of its properties. Its target-space action \cite{Grumiller:2005sq} is given by \eqref{ActionConclusion} with $8\pi G_2=1$ and the potentials
 \eq{
U(X) = -\frac{\rho}{\rho^2+2(1+\sqrt{1+\rho^2})}\,,\qquad V(X) = -2 b^2 \,\rho\,.
}{eq:EBH1}
Here the canonical dilaton $X$ is related to a new field $\rho$ by
\eq{
X=\rho+\arcsinh{\rho} \,,
}{eq:EBH2}
and the parameter $b$ is related to the level $k$ and $\alpha'$ by
%\begin{equation}\label{eq:EBH3}
$\alpha'\,b^{2} = 1/(k-2)$.
%\end{equation}
In order for the background following from \eqref{ActionConclusion}, \eqref{eq:EBH1} and \eqref{eq:EBH2} to be a solution of string theory it must satisfy the condition $D - 26 + 6\,\alpha'\,b^2 = 0$. Because the target space here is two-dimensional, $D=2$, requiring the correct central charge fixes the level at the critical value $k_{\rm crit} = 9/4$. Following \cite{Kazakov:2001pj}, we vary $k$ by allowing for additional matter fields that contribute to the total central charge, so that $k \in [2,\infty)$ is possible. The Witten BH arises in the limit $k\to\infty$. Since it is not possible to place an abrupt cut-off on the space-time fields in string theory we have to consider the limit $X_c\to\infty$ in the Helmholtz free energy \eqref{FreeEnergy}, 
\eq{
F^{ESBH}=-b\,\sqrt{1-\frac{2}{k}}\,\arcsinh{\sqrt{k(k-2)}}
}{eq:EBH10}
and its thermodynamical descendants. It is straightforward to show that \eqref{eq:EBH10} leads to a positive specific heat for any $k\in(2,\infty)$, and that it vanishes in the limit $k\to 2$ in accordance with the third law.

We comment now briefly on the inclusion of semi-classical corrections from fluctuations of massless matter fields on a given BH background. We therefore add to the classical action \eqref{eq:bulkaction} the Polyakov action,
\eq{ 
I_{\rm bulk}^{\rm semi} = I_{\rm bulk} + c \int_{\MM}\bns d^2x\sqrt{g}\left[\psi R + \frac12 (\nabla\psi)^2\right]\,,
}{eq:Pol}
where we have introduced an auxiliary field $\psi$ which fulfills the on-shell relation $\square\psi=R$, and the constant $c$ depends on the number and type of massless matter fields.  Obviously, the addition of \eqref{eq:Pol} requires a reconsideration of boundary issues. One possibility is to demand that $\psi$ is a function of $X$ \cite{Zaslavsky:1998ca}. %,Zaslavskii:2003cr}. 
Then the action \eqref{eq:Pol} reduces to a standard dilaton gravity action (we set $8\pi G_2=1$)
\eq{
I_{\rm bulk}^{\rm semi} = -\frac{1}{2} \int_{\MM}\bns d^2x\sqrt{g}\left[\hat{X}R-(U(X)+c (\psi'(X))^2)(\nabla X)^2-2V(X)\right]\,,
}{eq:redefaction}
upon introducing a redefined dilaton $\hat{X}=X-2\,c\,\psi(X)$. Therefore, as long as the assumption $\psi=\psi(X)$ is meaningful, the discussion of boundary terms in Section \ref{se:2} is still valid.  We note in this context that the large $X$ expansion of the exact string BH \eqref{eq:EBH1}, \eqref{eq:EBH2} can be interpreted as a semi-classical correction to the Witten BH, with $\rho$ playing the role of the unperturbed dilaton $X$ and $-\ln{(2\rho)/(2\,c)}$ playing the role of the auxiliary field $\psi$. This is consistent with the fact that the conformal factor $\psi$ scales logarithmically with the dilaton and concurs with semi-classical corrections \cite{Grumiller:2003mc} to the specific heat of the Witten BH, which also show qualitative agreement with the specific heat of the exact string BH. Another, more general, possibility is to treat $\psi$ as an independent field. In that case boundary issues have to be reconsidered. We expect  them to be relevant whenever the boundary term
\eq{
\int_{\dM}\bns dx\sqrt{\gamma} \,\psi n^\mu\partial_\mu\psi 
}{eq:scbt}
does not fall off sufficiently fast at the asymptotic boundary ($n^\mu$ is the outward pointing unit normal). Since $\psi$ typically scales logarithmically with $X$ this happens for all models where $w(X)$ grows linearly or faster than $X$. Interestingly, the Witten BH is precisely the limiting case where this issue is of relevance. More recently semi-classical corrections were considered in the context of large $AdS$ BHs \cite{Hemming:2007yq}. There the issue is complicated because the matter fields couple non-minimally to the dilaton. Since the results of \cite{Hemming:2007yq} agree with ours in the large $X$ limit only,\footnote{The counterterm, (4.5) in \cite{Hemming:2007yq}, for finite values of $r$ differs from \eqref{LCT}, which yields $\FF(r^2)\sim r\sqrt{1+r^2/\ell^2}=1/\ell\,(r^2 + \ell^2/2 + \dots)$.} it would be interesting to analyze the counterterms using the Hamilton-Jacobi method, possibly by adapting the strategy described and applied in \cite{deBoer:1999xf}. %, Martelli:2002sp, Larsen:2004kf, Batrachenko:2004fd}. 
This would also allow to reconsider the path integral quantization of 2D dilaton gravity with matter in the presence of boundaries \cite{Kummer:1998zs} %,Grumiller:2000ah,Fischer:2001vz,Grumiller:2002dm,Bergamin:2004us,Grumiller:2004yq,Bergamin:2005pg,Grumiller:2006ja} 
and to clarify the role of the Hamilton-Jacobi counterterm for observables beyond thermodynamical ones. 

For further applications and an outlook to future research we refer to the discussion in Section 7 of \cite{Grumiller:2007ju}.

\section*{Acknowledgments}

The content of this proceedings contribution was presented by one of us at the conference QFEXT07 in Leipzig, and DG would like to thank Michael Bordag as well as Boris, Irina, Mischa and Sascha Dobruskin for the kind hospitality.

%... support by LB ...
%
The work of DG is supported in part by funds provided by the U.S. Department of Energy (DoE) under the cooperative research agreement DEFG02-05ER41360. DG has been supported by the Marie Curie Fellowship MC-OIF 021421 of the European Commission under the Sixth EU Framework Programme for Research and Technological Development (FP6).
The research of RM was supported by DoE through grant DE-FG02-91ER
40688 Task A (Brown University) and Perimeter Institute for
Theoretical Physics. Research at Perimeter Institute is supported
through the government of Canada through Industry Canada and by the
province of Ontario through the Ministry of Research and Innovation.
%
%... support by RMe ...

\appendix

\section{Definitions of $w$ and $Q$}

The classical solutions of the equations of motion
\eq{
   X = X(r)\,, \qquad ds^2 = \xi(r) \,d\tau^2 + \frac{1}{\xi(r)}\,dr^2\,,
}{metric}
with
\eq{
  	\partial_r X =  e^{-Q(X)} \,, \qquad
	\xi(X) =  w(X) \, e^{Q(X)}\,\left( 1 - \frac{2\,M}{w(X)} \right)\,,  
}{eq:defs} 
are expressed in terms of two model-dependent functions,
\eq{
        Q(X) \defeq Q_0 \, + \int^{X} \bns d\tilde{X} \, U(\tilde{X})\,, \qquad
        w(X) \defeq w_0 -2 \, \int^{X} \bns d\tilde{X} \, V(\tilde{X}) \, e^{Q(\tilde{X})}\,.
}{QwDef}
Here $Q_0$ and $w_0$ are constants, and the integrals are evaluated at $X$. Notice that $w_0$ and the integration constant $M$ contribute to $\xi(X)$ in the same manner. Together, they represent a single parameter that has been partially incorporated into the definition of $w(X)$. By definition they transform as $w_0 \to e^{\Delta Q_0} w_0$ and $M \to e^{\Delta Q_0} M$ under the shift $Q_0 \to Q_0 + \Delta Q_0$. This ensures that the functions \eqref{eq:defs} transform homogeneously, allowing $Q_0$ to be absorbed into a rescaling of the coordinates. Therefore, the solution is parameterized by a single constant of integration. With an appropriate choice of $w_0$ we can restrict $M$ to take values in the range $M \geq 0$ for physical solutions. The function $w$ is invariant under dilaton dependent Weyl rescalings of the metric, whereas $Q$ transforms inhomogeneously. All classical solutions \eqref{metric} exhibit a Killing vector $\partial_\tau$. With Lorentzian signature each solution $X_h$ of $\xi(X)=0$ therefore leads to a Killing horizon. The Hawking temperature is given by the inverse periodicity in Euclidean time, $T=w'(X_h)/(4\pi)$.

\section*{References}

%\bibliographystyle{fullsort}
%\bibliography{review} 

\providecommand{\href}[2]{#2}\begingroup\raggedright\endgroup

\end{document}